\documentclass[manuscript]{aastex}

\shorttitle{Photoionization modeling of oxygen K absorption}
\shortauthors{Gatuzz et al.}

\begin{document}

\title{Photoionization modeling of oxygen K absorption in the interstellar medium:\\
       the Chandra grating spectra of XTE~J1817-330}

\author{E.~Gatuzz\altaffilmark{1},
        J.~Garc\'ia\altaffilmark{2,3},
        C.~Mendoza\altaffilmark{1,4},
        T.R.~Kallman\altaffilmark{3},
        M.~Witthoeft\altaffilmark{3},
        A.~Lohfink\altaffilmark{2},
        M.A.~Bautista\altaffilmark{5},
        P.~Palmeri\altaffilmark{6},
        \and P.~Quinet\altaffilmark{6,7}}

\altaffiltext{1}{Centro de F\'isica, Instituto Venezolano de Investigaciones
Cient\'ificas (IVIC), PO Box 20632, Caracas 1020A,  Venezuela
\email{egatuzz@ivic.gob.ve, claudio@ivic.gob.ve}}

\altaffiltext{2}{Department of Astronomy and Maryland Astronomy Center for
Theory and Computation, University of Maryland, College Park, MD 20742, USA
\email{javier@astro.umd.edu, alohfink@astro.umd.edu}}

\altaffiltext{3}{NASA Goddard Space Flight Center, Greenbelt, MD 20771, USA
\email{timothy.r.kallman@nasa.gov, michael.c.witthoeft@nasa.gov}}

\altaffiltext{4}{Centro Nacional de C\'alculo Cient\'ifico Universidad de Los Andes (CeCalCULA), Corporaci\'on Parque Tecnol\'ogico de M\'erida, M\'erida 5101, Venezuela}

\altaffiltext{5}{Department of Physics, Western Michigan University, Kalamazoo,
MI 49008, USA \email{manuel.bautista@wmich.edu}}

\altaffiltext{6}{Astrophysique et Spectroscopie, Universit\'e de Mons - UMONS,
B-7000 Mons, Belgium \email{palmeri@umons.ac.be, quinet@umons.ac.be}}

\altaffiltext{7}{IPNAS, Sart Tilman B15, Universit\'e de Li\`ege, B-4000 Li\`ege,
Belgium}

\begin{abstract}
We present detailed analyses of oxygen K absorption in the interstellar medium
(ISM) using four high-resolution {\em Chandra} spectra towards the X-ray low-mass
binary XTE J1817-330. The 11--25~\AA\ broadband is described with a simple absorption
model that takes into account the pileup effect and results in an estimate of the
hydrogen column density. The oxygen K-edge region (21--25~\AA) is fitted with the
physical {\tt warmabs} model, which is based on a photoionization model grid generated
with the {\sc xstar} code with the most up-to-date atomic database. This approach allows a
benchmark of the atomic data which involves wavelength shifts of both the K lines and
photoionization cross sections in order to fit the observed spectra accurately.
As a result we obtain: a column density of $N_{\rm H}=1.38\pm0.01\times 10^{21}$~cm$^{-2}$;
ionization parameter of $\log \xi=-2.70\pm0.023$; oxygen abundance of
$A_{\rm O}= 0.689^{+0.015}_{-0.010}$; and ionization fractions of \ion{O}{1}/O = 0.911,
\ion{O}{2}/O = 0.077, and \ion{O}{3}/O = 0.012 that are in good agreement with
previous studies. Since the oxygen abundance in {\tt warmabs} is given relative to
the solar standard of \citet{gre98}, a rescaling with the revision by \citet{asp09}
yields $A_{\rm O}=0.952^{+0.020}_{-0.013}$, a value close to solar that reinforces the new standard.
We identify several atomic absorption lines---K$\alpha$, K$\beta$, and K$\gamma$
in \ion{O}{1} and \ion{O}{2}; and K$\alpha$ in \ion{O}{3}, \ion{O}{6}, and
\ion{O}{7}---last two probably residing in the neighborhood of the source rather than
in the ISM. This is the first firm detection of oxygen K resonances with principal
quantum numbers $n>2$ associated to ISM cold absorption.
\end{abstract}



\section{Introduction}

X-ray studies of interstellar absorption have the potential to provide information that
is not accessible by other techniques. This includes the relative abundances of a wide
range of ion stages of interstellar cosmic elements (e.g. oxygen) and the abundances relative to H and He of elements with atomic number $Z\geq 6$.  X-ray absorption can also
provide signatures of the binding of these elements in molecules or solids;
the inner-shell electronic transitions are key diagnostics since the
ionization state or chemical binding shifts the line energies by a predictable amount.
The ongoing effort to explore and exploit this effect in interstellar studies involves
accumulation of synthetic spectra for inner-shell absorption of ions and neutrals,
based on both atomic calculation and laboratory experiment, in addition to testing them
against available observed interstellar X-ray absorption spectra.

As part of the effort to accumulate the required atomic data, the
atomic database of the {\sc xstar} photoionization modeling code \citep{bau01}
has been systematically improved in the past ten years in order to study the K lines
and edges in the high-quality X-ray astronomical spectra obtained from the {\it Chandra}
and {\it XMM-Newton} satellite-born observatories. This database includes K-vacancy
levels, wavelengths, $A$-values, radiative and Auger widths and high-energy photoionization
cross sections for complete isonuclear sequences with $Z\leq 30$ \citep[see][and references therein]{pal12}.

With this new database it has been possible, for instance, to estimate the model efficiency
of iron K-line emission and absorption in terms of the ionization parameter
and density \citep{kal04}. It has been shown that the centroid of the K$\alpha$ unresolved transition array, the
K$\beta$ energy, and the ratio of the K$\alpha_1$ to K$\alpha_2$ components are useful
diagnostics of the ionization parameter. It was also found that the many
strongly damped resonances below the K-ionization thresholds lead to edge smearing, which
can certainly hamper the astrophysical interpretation of the absorption features.
A synthetic spectrum has been modeled by \citet{kal09} for the black-hole X-ray transient
GRO~J1655-40 which constrains the odd-$Z$ element abundances ($11\leq Z\leq 27$)
and outflow parameters of the associated warm absorber. By analyzing the oxygen K-absorption
structure in the interstellar medium (ISM) from the {\em XMM-Newton} spectrum of the low-mass X-ray
binary Sco~X-1, \citet{gar11} reproduced both the K edge and \ion{O}{1} K$\alpha$ absorption line,
thus evaluating the impact of the atomic data on the interpretation of observations.
In the present report we continue this work with a more stringent benchmark of the
{\sc xstar} atomic database, made possible by {\em Chandra} spectra of a low-mass binary
displaying imprints of the ISM oxygen K absorption with higher resolution.

\citet{jue04} measured high-resolution spectra of the interstellar oxygen K-shell absorption edge from seven
X-ray binaries---Cygnus X-1, Cygnus X-2, 4U~1636-53, 4U~1735-44, GX~9+9, 4U~1543-624, and
4U~1820-30---using the {\it Chandra} High Energy Transmission Grating Spectrometer (HETGS).
The oxygen column density $N_{\rm O}$ was therein calculated using the optical depth at
$21.7$~\AA\ and the cross section at this wavelength assuming the oxygen abundance from \citet{wil00}.
With this value, the hydrogen column density $N_{\rm H}$ was then determined. The 1s--2p transitions
of \ion{O}{2} and \ion{O}{3} were identified, finding ISM abundances for \ion{O}{2} and \ion{O}{3}
relative to \ion{O}{1} of $\approx 0.1$ and $\leq 0.1$, respectively. This became the first estimate
of the \ion{O}{2}/\ion{O}{1} and \ion{O}{3}/\ion{O}{1} abundance ratios in the ISM.

By observing the oxygen K-shell edge toward galaxy clusters, \citet{bau06} showed that the ISM hydrogen X-ray column density was in close agreement with the $21$~cm radio value for columns less than approximately $0.5\times 10^{21}$~cm$^{-2}$. For higher column densities, the former was higher by as much as a factor of 2.5, indicating substantial absorption besides that due to neutral hydrogen (probably from clouds of molecular hydrogen). An average ISM oxygen abundance of 0.99 solar was found relative to the photospheric value ($8.66\pm 0.05$) by \citet{asp04} which implied a high gas to dust ratio.

\citet{pin10} reported an spectral model suggesting that, along the line of sight towards an X-ray source (GS~1826-238), the ISM was more complex than just simple neutral gas. Their gas model consisted of three components: cold gas ($T\approx 5.8{-}10\times 10^{3}$~K); warm ionized gas ($T\approx 1{-}6\times 10^{4}$~K);
and hot ionized gas ($T\approx 2\times 10^{6}$~K). It was also shown that the column densities
$N_{\rm H}$ of these three components span two orders of magnitude, where the cold gas contributed $\sim  90{-}95\%$, the warm gas $\sim 5{-}10\%$, and the hot gas $\sim 1\%$ to the total $N_{\rm H}^{\rm tot}$.
\citet{cos12} analyzed the ISM using the bright X-ray binary 4U 1820-30 source, concentrating on the mildly ionized cold gas where the model consisted of mildly ionized gas, dust, and mildly ionized outflowing gas. In the oxygen edge region, dust compounds were included which did not show features below 23.7~\AA. Finally, oxygen was found to be overabundant by a factor of 1.23 solar.

In order to model the oxygen K absorption features in the ISM with {\sc xstar}, we need a {\it Chandra}
spectrum with high counts. In the catalog of low-mass X-ray binaries by \citet{liu07}, we
found in XTE~J1817-330 a suitable source with an average count number of $\approx 8\times 10^{6}$ counts/s.
Discovered by the Rossi X-ray Timing Experiment \citep{rem06}, XTE~J1817-330 is a low-mass X-ray binary
displaying a primary component which has an X-ray spectrum and variability behavior which are consistent with
its identification as a black hole \citep{sal07}. Its location, $l=359.81$ and $b=-7.99$ in Galactic coordinates,
implies a line of sight in the Galactic center direction, a region that dominates the total mass of the Galaxy
\citep{Sta12} providing a good opportunity to study the ISM absorption features. This source has been observed
in different spectral ranges, namely the optical \citep{tor06}, near infrared \citep{dav06}, radio \citep{rup06}, and
X-ray \citep{roy11}. \citet{sal07} fitted its X-ray spectrum finding an absorbing hydrogen column of
$1.55\times 10^{21}$~cm$^{-2}$ and identifying the following oxygen absorption lines: \ion{O}{1}
K$\alpha (23.52$~\AA); \ion{O}{2} K$\alpha (23.35$~\AA); \ion{O}{3} K$\alpha (23.13$~\AA); \ion{O}{1}
K$\beta (22.91$~\AA); and \ion{O}{7} K$\beta (21.60$~\AA). \citet{gie09} estimated a hydrogen column density
of $N_{\rm H}=1.08\pm 0.05\times 10^{21}$~cm$^{-2}$, and \citet{roy11} investigated the timing
characteristics in the spectra using a column of $N_{\rm H}=1.2\times 10^{21}$~cm$^{-2}$.

The outline of the present report is as follows. In Section~2, we describe the {\sc xstar} {\tt warmabs} physical model which is used to obtain the final spectrum fits, in particular the details concerning its atomic database. The data reduction aspects of the observations, e.g. the pileup effect, are reviewed in Section~3 while the results of the broadband fits and resolved absorption features are given in Sections~4--5. Section~6 is dedicated to comparisons of the hydrogen column density that allow an estimate of its reliability. Section~7 describes the oxygen column densities derived from our fits.  Finally, a discussion and conclusions are presented in Section~8.


\section{XSTAR/warmabs}

{\sc xstar}\footnote{http://heasarc.nasa.gov/lheasoft/xstar/xstar.html} is a code package
designed to determine the physical conditions in partially ionized gases. It calculates
the temperature, ionization, excitation, absorption, and emission due to neutral and
all ionized species of elements with atomic number $Z\leq30$. As many relevant physical processes
as possible are taken into account, assuming a time stationary balance among them and a Maxwellian
electron velocity distribution. Important processes for many problems relevant to X-ray
astronomy include photoionization, electron impact collisional ionization and excitation,
and radiative and dielectronic recombination. More complete descriptions of {\sc xstar}
and its atomic database are given by \citet{kal01} and \citet{bau01}. In this work we employ the {\tt warmabs} utility which allows for the calculation of {\sc xstar} models within the
widely used X-ray spectral fitting package
{\sc xspec}\footnote{http://heasarc.gsfc.nasa.gov/xanadu/xspec/}.
It utilizes precalculated tabulations of the ion fractions and level populations in
order to calculate synthetic emission and transmission spectra. It assumes that the gas
responsible for emission or absorption has a uniform ionization and temperature throughout,
although it is possible to superimpose two or more components and simulate a non-uniform situation.


\subsection{Atomic data}

The K photoabsorption data sets for oxygen ions with electron numbers $2\leq N \leq 8$ contained
in the {\sc xstar} atomic database were generated by \citet{gar05}. Energies for both valence
and K-vacancy fine-structure levels, transition wavelengths, $gf$-values, and radiative and
Auger widths were computed with the atomic structure packages {\sc autostructure}
\citep{eis74, bad86, bad97} and {\sc hfr} \citep{cow81}. Electron correlation effects were taken
into account by means of configuration-interaction expansions within the $n\leq 3$ complexes,
and relativistic corrections were included with a Breit--Pauli Hamiltonian. The accuracy of the
theoretical level energies and wavelengths does not generally match that attained in experiments
(see Table~\ref{ebit}), and the scanty availability of measurements for the K shell of O ions with
$N> 3$ limits the implementation of term-energy corrections and wavelength shifts that can be
introduced in order to compensate for the theoretical shortcomings. Furthermore, as also shown in Table~\ref{ebit},
experimental misassignments, such as that quoted by \citet{gu05} for \ion{O}{6}, are not uncommon. (The
discrepancy in \ion{O}{3} is believed to be a typo.)

High-energy photoionization cross sections in the K-edge region for oxygen ions were obtained
by \citet{gar05} with {\sc bprm}, the Breit--Pauli $R$-matrix suite of codes \citep{sco80, sco82, ber87, sea87}.
A conspicuous effect in the resonance structure of these cross sections that must be treated in detail
is Auger damping. It is the result of the dominant spectator KLL Auger decay channels of a photoexcited
K-vacancy state, say,
\begin{eqnarray}
[\rm{1s}]\rm{2p}^\mu n\rm{p}& \rightarrow & \rm{2p}^{\mu-2}n\rm{p}+\rm{e}^-\\
                            & \rightarrow & [\rm{2s}]\rm{2p}^{\mu-1}n\rm{p}+\rm{e}^-\\
                            & \rightarrow & [\rm{2s}]^2\rm{2p}^{\mu}n\rm{p}+\rm{e}^-
\end{eqnarray}
over the participator Auger KL$n$ channels
\begin{eqnarray}
[\rm{1s}]\rm{2p}^\mu n\rm{p}& \rightarrow & \rm{2p}^{\mu-1}+\rm{e}^-\\
                            & \rightarrow & [\rm{2s}]\rm{2p}^{\mu}+\rm{e}^- \ ,
\end{eqnarray}
which leads to edge smearing by resonances with symmetric profiles of nearly constant width. They are
handled within the $R$-matrix package by means of an optical potential \citep{gor99, gor00}. Again,
the accuracy of the resonance energy positions is limited by the absence of measurements which would
enable empirical adjustments of the many K-vacancy target thresholds (i.e. the series limits) in the
close-coupling expansion.


\section{Observations and data reduction}

In Table \ref{tab1} we list the specifications of the four observations of XTE J1817-330 obtained by {\it Chandra}
using the High Energy Transmission Grating Spectrometer (HETGS) in combination with the
Advanced CCD Imaging Spectrometer (ACIS). The oxygen edge is accessible with the
HETGS exclusively via the Medium Energy Gratings (MEG). Given the large flux of this source ($F = 1350~\mu$J, integrated in the 2--10~keV energy range), these observations were
taken in continuous clocking mode which significantly increases the temporal resolution
in order to minimize the {\it pileup} effect \citep{cac08}. The pileup is an inherent feature of CCD
detectors, such as those of the ACIS instrument, which occurs when two or more photons are
detected as a single event possibly causing a deformation in the level and shape of the
continuum \citep{mil06a}. It is stronger in the MEG than in the HEG due to
the lower dispersion and higher effective area of the former\footnote{http://cxc.harvard.edu/ciao/ahelp/acis\_pileup.html}.
Although the pileup is usually not an issue near the oxygen edge, it may be for the continuum
which is used to establish the hydrogen column density by fitting the 11--25~\AA\ interval; hence, we consider
the pileup effect in this range by applying the correction model {\tt simple\_gpile2.sl}
\citep{han09} before spectral fitting. Among the most popular computer packages for fitting X-ray
data are {\sc xspec} and {\sc isis}, and we have used the latter
(version 1.6.2\footnote{http://space.mit.edu/cxc/isis/}) to include the pileup model.

Even though the spatial resolution is reduced to one dimension when the continuous clocking mode
is used, there is essentially no difference with the timed exposure mode in the background
extraction or in the data reduction process. Hence, we have reduced the data sets using the
standard CIAO threads\footnote{http://cxc.harvard.edu/ciao4.4/threads/gspec.html}.
In some cases, the zero-order data were not telemetered; for these observations
the zero-order position was then estimated by finding the intersection of the grating arms
using the {\tt findzo} algorithm\footnote{http://space.mit.edu/cxc/analysis/findzo/}.

\begin{figure}
\epsscale{1.0}
\plotone{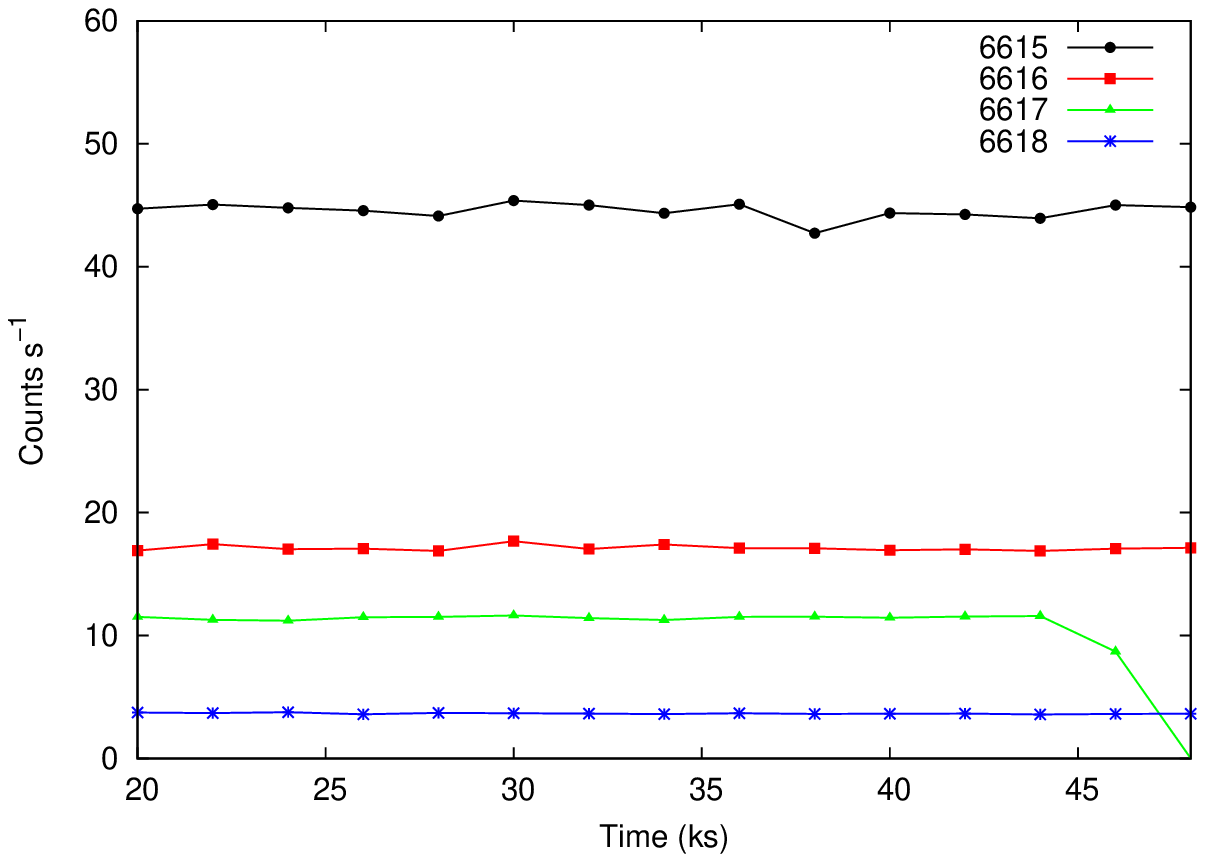}
\caption{Light curves of the four observations, ObsID 6615, 6616, 6617, and 6618, in units of counts/s. The approximately constant average counts/s that prevails is an indication of a low variability degree.}
\label{lightcurve}
\end{figure}

Figure \ref{lightcurve} shows the light curves of the four observations ObsID 6615, 6616, 6617, and 6618 in units of counts/s. It may be seen that the average counts/s is approximately constant for all, and therefore, we use the averaged spectra for this work.
We have carried out the following spectral fitting procedure: firstly, each spectrum was rebinned with
a 0.1~\AA\ bin in order to fit the continuum in the 11--25~\AA\ wavelength range so as to find the
hydrogen column density using the {\tt TBnew} model; then, the default spectral resolution was restored,
and the oxygen edge region (21--24~\AA) was analyzed with the {\tt warmabs} model freezing the
column density at the value obtained in the preceding step.

\section{Broadband fit}\label{sbro}

Following \citet{yao06}, we rebinned each spectrum with a 0.1~\AA\ bin size throughout the
spectral range to decrease the number of absorption lines and improve the broadband fit. We fitted
the four observations simultaneously in the interval 11--25~\AA\ using the {\tt simplegpile2(TBnew(powerlaw))}
model, where {\tt simplegpile2} is a convolution model to account for pileup effects and {\tt TBnew}
is an X-ray absorption model that includes abundances for elements from H to Ni \citep{wil00}.
Also we added Gaussians to fit the remaining absorption lines after rebinning. We used abundances
specified in \citet{wil00} and cross sections by \citet{ver96}. The O, Ne, and Fe abundances were
handled as free parameters, and for the analysis $\chi^{2}$ statistics was employed. We fixed the
absorption parameters in all the observations, and varied the power-law, pileup, and Gaussian
parameters for each case.

\begin{figure}
\epsscale{1.0}
\plotone{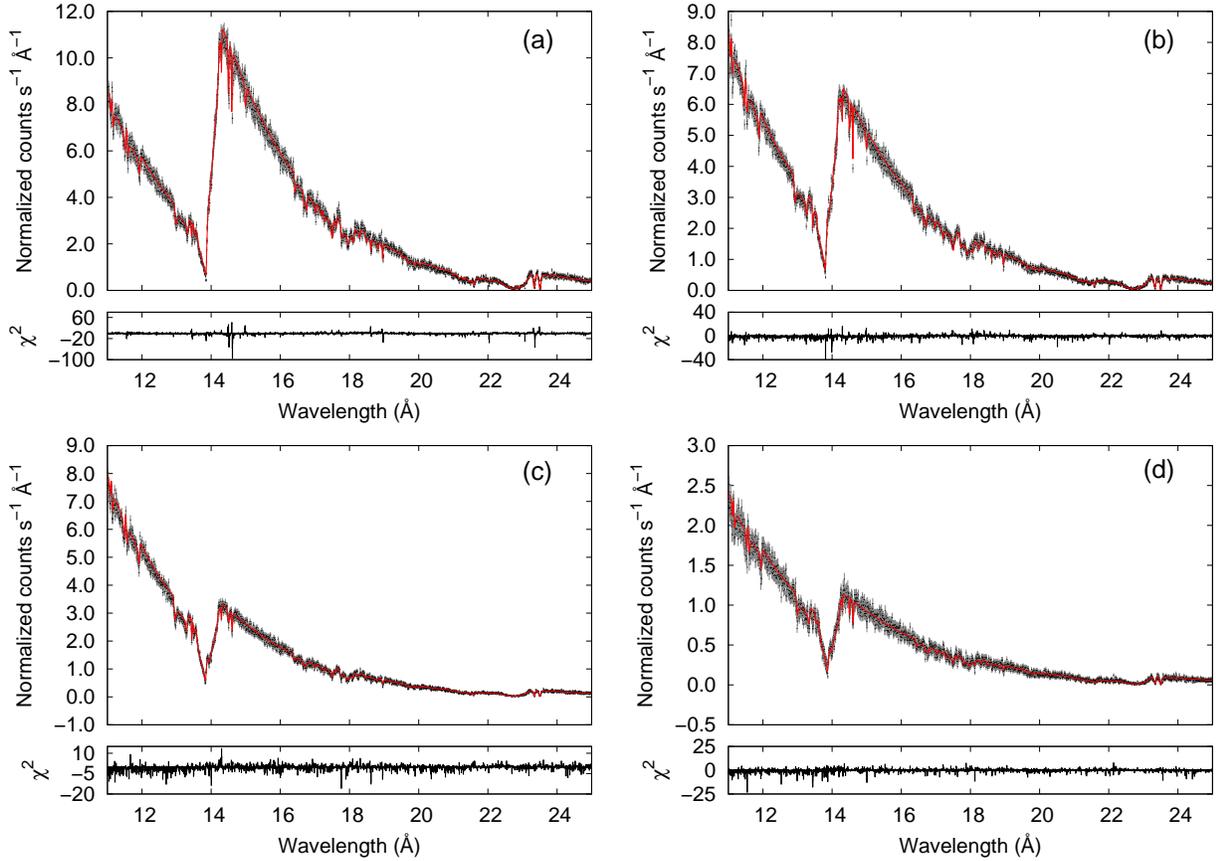}
\caption{{\it Chandra} MEG spectra of the X-ray binary XTE~J1817-330 simultaneously fitted in the 11--25~\AA\ region using the {\tt TBnew} model. (a) ObsID 6615. (b) ObsID 6616. (c) ObsID 6617. (d) ObsID 6618. See Table~\ref{tab1} for further details pertaining these observations. In each panel, the black data points are the observations while the solid red lines correspond to the best-fit models. Residuals are shown as histograms for each case in units of $\chi^2$. Note the large residuals in the oxygen K-edge region, particularly in the neighborhood of the \ion{O}{1} and \ion{O}{2} K$\alpha$ lines at $\approx 23.50$~\AA\ and $\approx 23.35$~\AA, respectively.}
\label{ftbnew}
\end{figure}

Figure \ref{ftbnew} shows the fit in the interval 11--25~\AA\ using the {\tt simplegpile2(TBnew(powerlaw))}
model. Although the fit is carried out simultaneously for the four observations, they are plotted
separately to improve clarity. Panels (a), (b), (c), and (d) correspond
to observations ObsID 6615, 6616, 6617, and 6618, respectively (see Table~\ref{tab1}).
In each panel, black data points correspond to the observation while the
solid red lines represent the best-fit model for each case, and the base plots show the fit residuals
in units of $\chi^2$. In each spectrum the K edges of Ne and O are respectively located at
$\approx 14.3$~\AA\ and $\approx 23.2$~\AA. The \ion{O}{1} and \ion{O}{2} 1s--2p absorption lines
at 23.5~\AA\ and 23.35~\AA\ are also clearly observed. We improved the fit by including Gaussians
to model the outstanding absorption lines: \ion{Ne}{9} K$\alpha$
($\approx 13.43$~\AA); \ion{Ne}{3} K$\alpha$ ($\approx 14.50$~\AA); \ion{Ne}{2} K$\alpha$
($\approx 14.60$~\AA); \ion{O}{8} Ly$\alpha$ ($\approx 18.95$~\AA); \ion{O}{7} K$\alpha$
($\approx 21.58$~\AA); and \ion{O}{2} K$\alpha$ ($\approx 23.34$~\AA). Residuals
indicate that the model fits the data satisfactorily with deviations near the Ne edge
($\approx 14.3$~\AA) due to background (i.e. clocking the whole chip) or an unaccounted
hot pixel. We also find high residuals in the oxygen absorption region; in particular,
the \ion{O}{1} K${\alpha}$ and \ion{O}{2} K${\alpha}$ are not well modeled.

Fit results for the individual free abundances, column density, power law,
and pileup are presented in Table~\ref{tab2}. The fit statistics give a
reduced chi-square of $\chi ^{2} = 1.208$. Differences in the power-law
parameters are due to the count numbers in each observation. We obtain
abundances for O, Ne, and Fe higher than solar \citep{wil00}, especially for Ne (factor of 2),
and a column density of $N_{\rm H}=1.66^{+0.03}_{-0.04}\times 10^{21}$ cm$^{-2}$
which is somewhat larger than those estimated in other spectral fits and by
measurements of the $21$~cm line. This will be further discussed in Section~\ref{scol}.

The nonlinear pileup convolution model {\tt simple\_gpile2.sl} exponentially
reduces the predicted count rate $R_{j}(\lambda)$ (in units of counts/s/\AA)
according to
\begin{equation}
\label{eq1}
R(\lambda)=R(\lambda)\exp(-\beta R_{\rm tot}(\lambda))
\end{equation}
where $R_{\rm tot}(\lambda)$ is the total spectral count rate
and $\beta$ is treated as a fit parameter. The model is only applicable to
first order grating spectra although it employs information from the
second and third order grating spectra\footnote{http://cxc.harvard.edu/ciao/ahelp/acis\_pileup.html}.
With this model we obtained an average $\beta= 0.050$ according to which the
spectra have a pileup degree greater than 25$\%$ for $14\leq\lambda\leq 16$~\AA\
with the highest value of $p=42\%$ at $\approx 14.3$~\AA\ for the observation
with the largest number of counts (ObsID 6615). This value shows the need to include
the pileup effect in the 11--25~\AA\ region to obtain a good fit. However,
for all the observations, we have a pileup degree lower than 5\% in the
21--24~\AA\ wavelength range which corresponds to the oxygen absorption region
and, therefore, its effects were ignored.

\section{Absorption features}\label{secGauss}

\begin{figure}
\epsscale{1.0}
\plotone{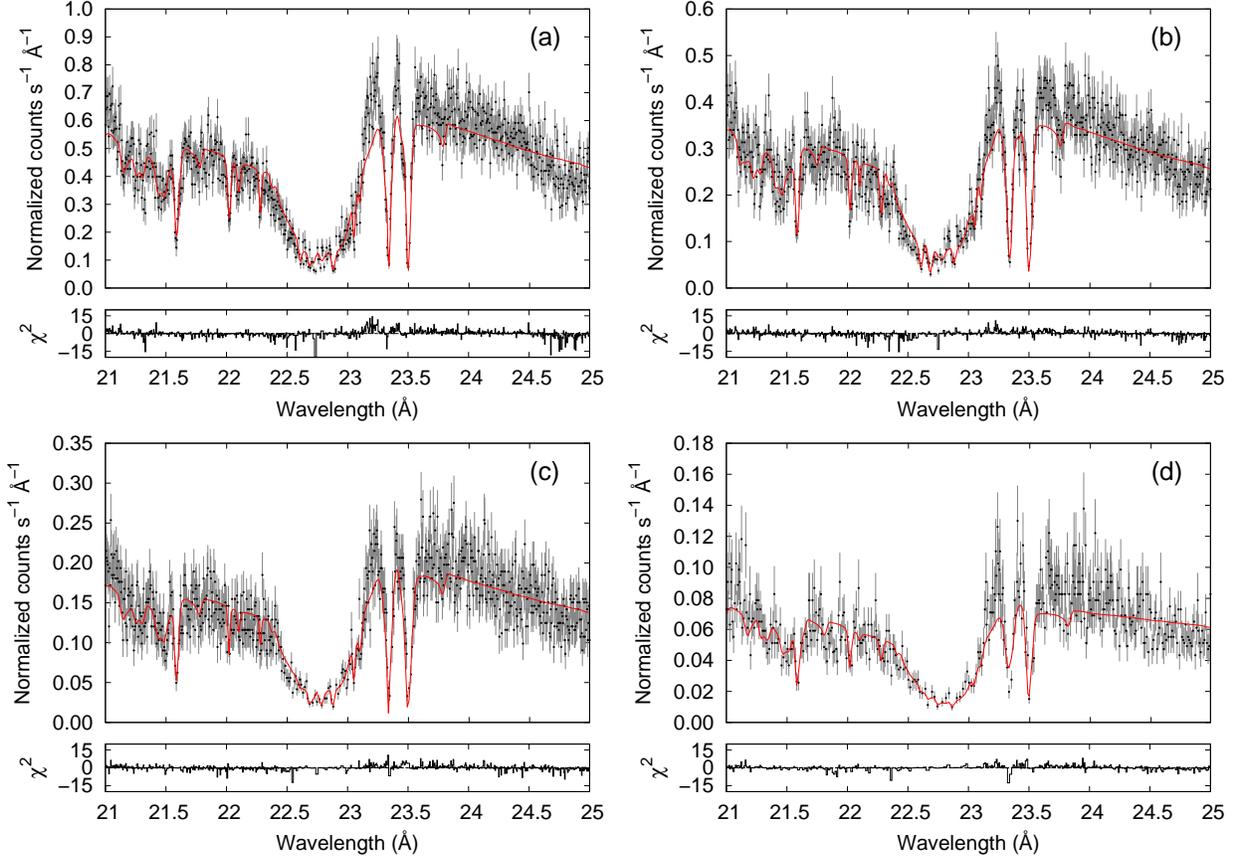}
\caption{Spectral fit of the {\it Chandra} MEG observations of XTE~J1817-330 in the
oxygen absorption region (21--25~\AA) using a simple functional model (power law and
several Gaussian profiles). (a) ObsID 6615. (b) ObsID 6616.
(c) ObsID 6617. (d) ObsID 6618. The observed absorption features are interpreted as:
the neutral O K-edge ($\approx 23.1$~\AA); the K$\alpha$, K$\beta$, and K$\gamma$ lines of
\ion{O}{1} ($\approx 23.50$~\AA, $\approx 22.88$~\AA, and  $\approx 22.68$~\AA, respectively);
the K$\alpha$, K$\beta$, and K$\gamma$ lines of \ion{O}{2} ($\approx 23.35$~\AA, $\approx 22.28$~\AA,
and  $\approx 22.10$~\AA, respectively); and the K$\alpha$ lines of \ion{O}{3}, \ion{O}{6}, and \ion{O}{7}
($\approx 23.10$~\AA, $\approx 22.02$~\AA, and $\approx 21.58$~\AA, respectively).
See Table~\ref{tab4} for more details.}
\label{fgauss}
\end{figure}

In order to identify all the absorption features present in the 21--25~\AA\
wavelength region, we first applied a functional model consisting of a power law
for the continuum and Gaussian profiles to describe the absorption lines.
Figure~\ref{fgauss} shows the spectral fit of the four {\it Chandra} MEG observations
of XTE~J1817-330 in the oxygen absorption region. Although shown in separate panels
for clarity, the fit was performed simultaneously for the four observations. Nevertheless,
the power-law parameters (photon index and normalization) are allowed to vary independently
for each data set. Several absorption features are clearly observed, the most outstanding
being the \ion{O}{1} and \ion{O}{2} K${\alpha}$ lines. The base plot in each panel shows
in units of $\chi^2$ the fit residuals as a histogram. High residuals are found around
24.8~\AA, and the model underestimates the number of photons around 23.2~\AA.

Table~\ref{tab4} shows a list of the lines obtained with the functional fit including the centroid wavelengths derived from the Gaussian profile parameters. It also lists the wavelengths for the lines measured by \citet{sal07}.
Note that our wavelengths are within the error bars of the values reported by
these authors, which are much larger given the lower spectral resolution of the
{\it XMM-Newton} instruments. We obtain a reduced chi-square of $\chi ^{2}=1.434$.
The \ion{O}{1} K${\alpha}$ is found at $23.502\pm 0.001$~\AA\ which has been previously
declared of ISM origin \citep{sal07}. Furthermore, we have also detected four other
\ion{O}{1} absorption lines corresponding to resonances with higher principal quantum number $n$:
two pairs of K$\beta$ ($n=3$) and K$\gamma$ ($n=4$) resonances associated to the $^4{\rm P}$ and
$^2{\rm P}$ K-hole core states. These are found at $22.884\pm 0.004$~\AA, $22.790 \pm 0.001$~\AA,
$22.686 \pm 0.004$~\AA, and $22.609 \pm 0.007$~\AA, respectively. In this set, only
the first K$\beta$ (belonging to the $^4{\rm P}$ core state) has been previously detected at
$22.91 \pm 0.03$~\AA\ in the {\it XMM-Newton} observation analyzed by \citet{sal07}.
This is then the first firm detection of oxygen K resonances with $n>2$ associated to ISM cold absorption.

The line at $23.358 \pm 0.002$~\AA\ is also of ISM origin \citep{jue04} and
corresponds to \ion{O}{2} K$\alpha$. As in the case of \ion{O}{1}, resonances with $n>2$
were also detected for \ion{O}{2}. These were found at $22.280 \pm 0.003$~\AA\
and $22.101 \pm 0.005$~\AA\, corresponding to K$\beta$ and K$\gamma$,
respectively. Although the \ion{O}{3} K$\alpha$ resonance is a triplet, only two absorption
features were detected at $23.104 \pm 0.005$~\AA\ and $23.054 \pm 0.001$~\AA\, while
\cite{sal07} only found one line at $23.13 \pm 0.09$~\AA\ identified as \ion{O}{3}
K$\alpha$. Finally, we have also located K$\alpha$ transitions for both
\ion{O}{6} ($22.022 \pm 0.003$~\AA) and \ion{O}{7} ($21.589\pm 0.003$~\AA), from which
only the strongest of the two (\ion{O}{7}) was reported by \citet{sal07} at
$21.609 \pm 0.06$~\AA.

The functional fit described above illustrates the complexity of oxygen K absorption.
There are (at least) two different regimes: a cold component of mostly neutral gas and a
hot component with highly ionized gas. Previous studies have suggested a multiple-phase
ISM; however, it is not clear if the two high-ionization lines from \ion{O}{6}
and \ion{O}{7} are intrinsic to the source or in fact arise in the ISM. Thus we
direct our study to the cold phase. Furthermore, since cold absorption is
not exclusively due to neutral oxygen, this means that a model such as {\tt TBnew} will
not completely represent the observed features. We have then simultaneously fitted
the four exposures in the 21--25~\AA\ wavelength range using the {\tt powerlaw*warmabs} physical model.
The advantage of using {\tt warmabs} is that it contains the most recent atomic
data for all the ions in the oxygen isonuclear sequence, thus enabling the fit
of all the absorption lines from \ion{O}{1}, \ion{O}{2}, and \ion{O}{3}. While implementing
this model, the column density is held fixed at the value obtained in the broadband
simultaneous fit ($N_{\mathrm H}=1.66 \times 10^{21}$~cm$^{-2}$). The ionization parameter
and the oxygen abundance are treated as free parameters but the same for all observations, while
the abundances of all the other chemical elements are held fixed at their solar values.
The quantities describing the power law (photon index and normalization) are also taken as
free parameters but independently for each observation.

\begin{figure}
\epsscale{1.0}
\plotone{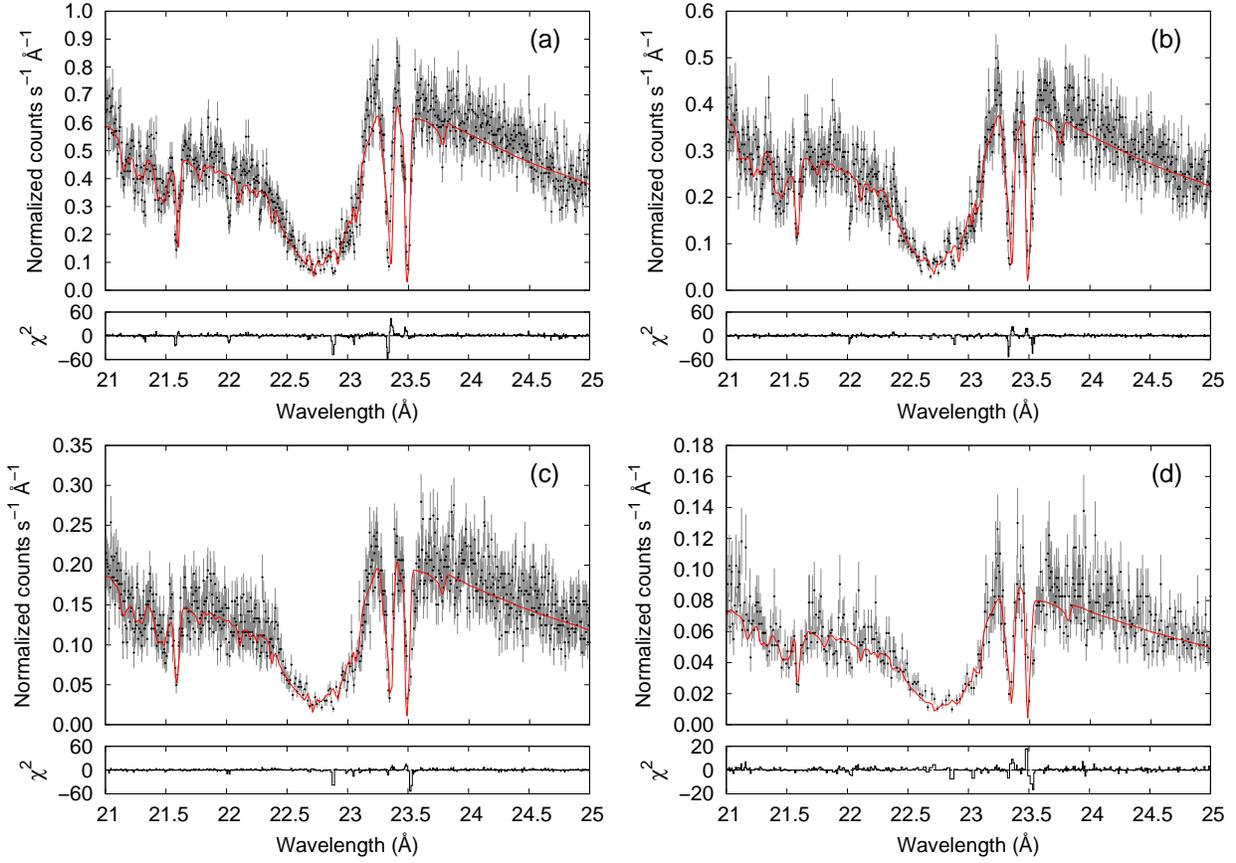}
\caption{Spectral fit of the {\it Chandra} MEG observations of XTE~J1817-330 in the
oxygen absorption region (21--25~\AA) using a {\tt powerlaw*warmabs} physical model.
(a) ObsID 6615. (b) ObsID 6616. (c) ObsID 6617.(d) ObsID 6618.}
\label{foldwarmabs}
\end{figure}

Figure \ref{foldwarmabs} shows the results of this fit. As before, we present for clarity the spectra
with the best-fit model in separate panels although the fit is performed simultaneously.
Since most of the relevant absorption features in the spectra are due to neutral, singly, and
doubly ionized O ions, the ionization parameter that describes the fit tends to
be somewhat low: log~$\xi=-2.729$. In this case, the {\tt warmabs} model does not show any absorption
due to \ion{O}{7}; therefore, this line is still represented with a Gaussian profile.
We have found large residuals around most of the oxygen absorption features, the
most prominent being around the \ion{O}{1} and \ion{O}{2} K$\alpha$ lines. The fit statistics
gives a poor reduced chi-square of $\chi^2=1.769$ which forced us to revise the atomic database.

\begin{figure}
\epsscale{1.0}
\plotone{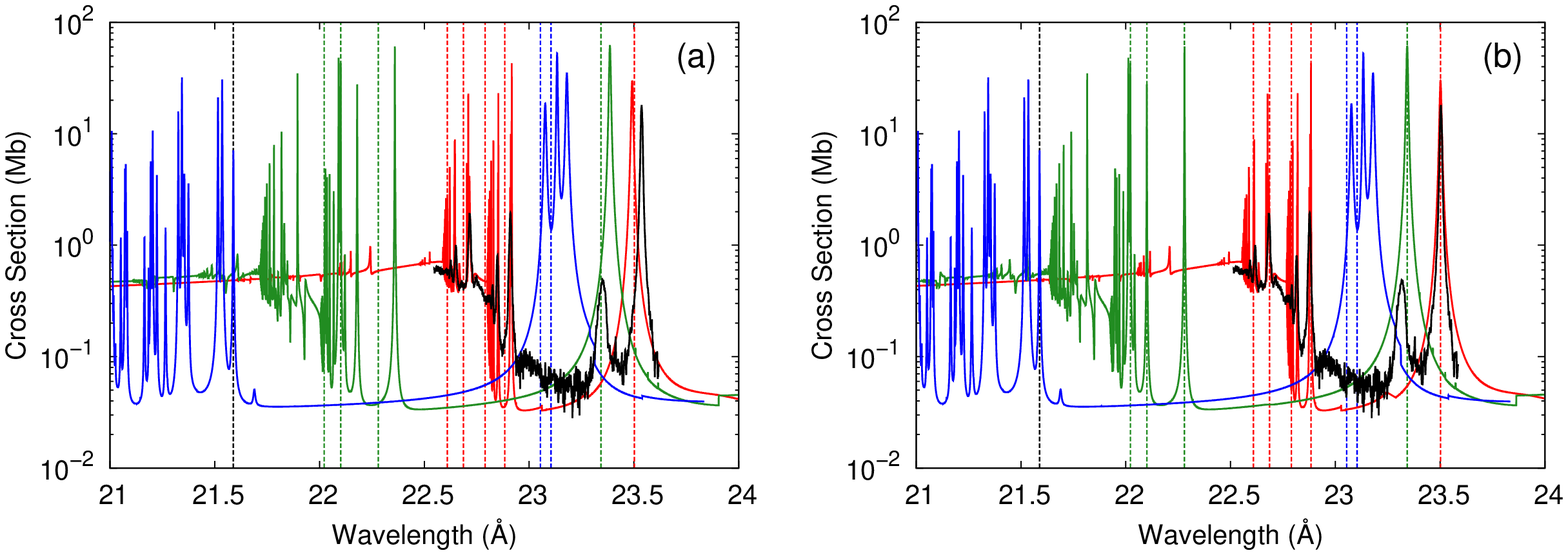}
\caption{Comparison of the theoretical photoabsorption cross sections for \ion{O}{1} (red),
\ion{O}{2} (green), and \ion{O}{3} (blue) computed by \citet{gar05} which are implemented
in the {\tt warmabs} model. The black solid line is the laboratory measurement by \cite{sto97}
showing peaks of \ion{O}{1} and O$_2$. The vertical dashed lines are placed at the wavelengths
of the observed absorption lines (see Table~\ref{tab4}). Panel (a) displays the original
cross sections while panel (b) shows the same curves after the wavelength shifts are applied
to both \ion{O}{1} and \ion{O}{2} (see text for details).}
\label{fxs}
\end{figure}

A close examination of the resonance positions from the photoabsorption cross sections
used in {\tt warmabs} in comparison with those obtained from the spectral fits with Gaussian profiles reveals
the inaccuracies of the atomic data. Figure~\ref{fxs}a shows with solid colored lines the atomic cross sections
used in {\tt warmabs} for \ion{O}{1}, \ion{O}{2}, and \ion{O}{3} \citep{gar05}.
Vertical dashed lines are placed at the positions of the absorption features found
with the functional fit described above (see Table~\ref{tab4}). The comparison shows that not
only the positions of the \ion{O}{1} and \ion{O}{2} K$\alpha$ resonances are displaced
with respect to the observed data but also those with $n>2$. The solid black line represents
the experimental \ion{O}{1} photoabsorption cross section measured by
\cite{sto97} (unfortunately, there are no laboratory measurements available for the other
species) which displays a similar wavelength shift with respect to the observed resonance positions.
This wavelength offset has been previously discussed by \cite{jue04} based on the K$\alpha$
position observed in the {\it Chandra} spectra from several sources, all of which match
the position we have found using Gaussian profiles. For \ion{O}{1}, the comparison
of the theoretical cross section with the laboratory measurement shows that, except the K$\alpha$
peak, all the other resonances agree very well. This implies that the relative position of the
K$\alpha$ with respect to the other resonances also needs to be adjusted. This is
also the case for \ion{O}{2} since the wavelength difference of the theoretical and observed K$\alpha$
positions is not the same as those for the rest of the resonances of the same ion. Consequently,
we have decided on the one hand to adjust the positions of the K$\alpha$ resonances (which are treated as lines in {\tt warmabs}), and on the other, to shift the whole cross sections for both species so as to obtain the best possible agreement with the observed lines. The new positions for the \ion{O}{1} and \ion{O}{2} K$\alpha$ lines are $23.502$~\AA\ and $23.343$~\AA\ while the cross sections were shifted towards shorter wavelengths by
$0.033$~\AA\ and $0.079$~\AA, respectively. No correction was applied to the \ion{O}{3}
curve since the observed lines are weaker, in addition to the fact that in this case the
K$\alpha$ resonance is a triplet. Figure~\ref{fxs}b shows the atomic absorption cross
sections after the wavelength corrections are applied. In other words, the
{\it Chandra} observed line positions suggest not only that the theoretical resonance positions from \cite{gar05} must be adjusted, but also that the wavelength scale of the only
available experiment for \ion{O}{1} \citep{sto97} needs to be shifted by $33$~m\AA.

Possible effects due to the instrument wavelength calibration, model uncertainties, and Doppler
shifts due to motion of the gas in the ISM are important for accurate line identification.
The instrumental resolution for the first-order spectra from MEG is $\Delta \lambda \sim 23$~m\AA,
which is better than the largest shift required for our theoretical cross sections ($33$~m\AA).
In fact, according to \cite{gar05}, the uncertainties in the theoretical cross sections for
O ions with electron occupancies $N\leq 4$ were estimated to be $\sim 50$~m\AA, comparable with
the wavelength uncertainties in the laboratory measurements of \cite{sto97}.
Assuming a velocity dispersion for the ISM of $\leq 200$ km s$^{-1}$ \citep{jue04}, the largest
wavelength shift due to Doppler effects for the \ion{O}{1} K$\alpha$ is $\Delta \lambda =14$~m\AA,
which is smaller than the instrumental resolution. Hence, we can safely ignore the effects of ISM
gas motion in the light of sight.

\begin{figure}
\epsscale{1.0}
\plotone{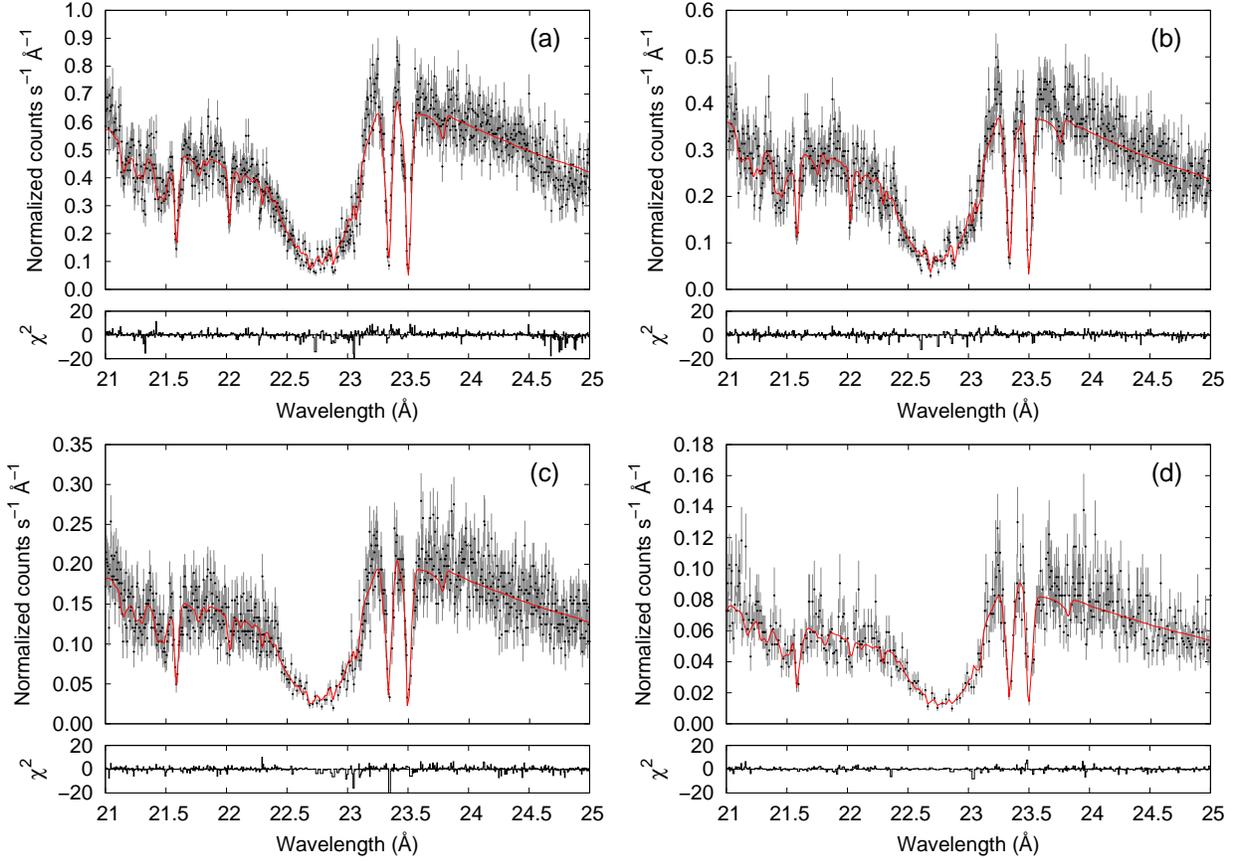}
\caption{Spectral fit of the {\it Chandra} MEG observations of XTE~J1817-330 in the
oxygen absorption region (21--25~\AA) using a {\tt powerlaw*warmabs} physical model,
after the wavelength corrections in the \ion{O}{1} and \ion{O}{2} atomic cross
sections are introduced. (a) ObsID 6615. (b) ObsID 6616. (c) ObsID 6617. (d) ObsID 6618.}
\label{fnewwarmabs}
\end{figure}

Figure~\ref{fnewwarmabs} shows the fit results with {\tt warmabs} after
the theoretical line positions have been corrected, where the residuals
are now significantly reduced and more evenly distributed. This is reflected in
better fit statistics (reduced $\chi^2=1.245$) and a well-constrained
ionization parameter ($\log\xi=-2.699\pm0.023$) which is consistent
with the low-ionization oxygen lines observed. As before, the \ion{O}{7} K$\alpha$
resonance at $21.589$~\AA\ is not well represented with this model and must be
fitted with a Gaussian profile.
We have also noticed that the absorption feature at $22.022$~\AA\ is not entirely due
to the \ion{O}{3} K$\delta$ resonance and is thus fitted with a Gaussian
profile, confirming a blend of both \ion{O}{3} K$\delta$ and
\ion{O}{6} K$\alpha$. In this final analysis, we also treated the column density value
as free parameter in order to increase the accuracy of the fit. We find a column
density of $N_{\mathrm H}=1.38\pm 0.1 \times 10^{21}$~cm$^{-2}$ that is lower than the value
previously mentioned in connection with the broadband fit (see Section~4). The oxygen abundance
obtained with the {\tt warmabs} model is $A_{\mathrm O}= 0.689^{+0.015}_{-0.010}$
relative to solar, also lower than the value for the best fit with {\tt TBnew}
($A_{\mathrm O}=1.178\pm 0.22$). All the best-fit parameters of the {\tt warmabs}
are summarized in Table~\ref{tab5}.


\section{Hydrogen column density}\label{scol}

Table~\ref{tab3} shows a comparison of the hydrogen column density obtained in this
work with previous estimates including measurements using the $21$~cm line.
By means of the simultaneous broadband fit described in Section~\ref{sbro}, we have obtained
a column density of $N_{\rm H}=1.66^{+0.03}_{-0.04}\times 10^{21}$ cm$^{-2}$
which is somewhat larger than those quoted in previous spectral fits.
\cite{mil06b} reported a hydrogen column density of $N_{\rm H}=8.8{-}9.7\times 10^{20}$~cm$^{-2}$
using a 50~ksec {\it Chandra} spectrum, a value smaller than those obtained from Galactic
\ion{H}{1} surveys, e.g. $1.58\times 10^{21}$~cm$^{-2}$ \citep{dic90}
and $1.39\times 10^{21}$~cm$^{-2}$ \citep{kal05}. Using the equivalent widths of
several interstellar bands in the optical spectrum, \cite{tor06} found the column
density to be in the range $N_{\rm H}=1{-}3\times 10^{21}$ cm$^{-2}$. Moreover,
both \citet{ryk07} and \citet{gie08} gathered acceptable fits of several
{\it Swift} XRT spectra with a fixed Galactic column density of
$N_{\rm H}=1.2\times 10^{21}$ cm$^{-2}$.
\cite{gie09} used the same XRT spectra combined with data from UVOT (The Ultraviolet and Optical
Telescope) to derive a column density of $N_{\rm H}=1.08\pm 0.05\times 10^{21}$ cm$^{-2}$
from broadband spectral fits.

The agreement between {\tt TBnew} and \cite{sal07} is around 7\% despite the fact that their
fits include data from {\it XMM-Newton} EPIC-Pn ($0.6{-}10.0$~keV), RGS1 ($0.3{-}2.0$~keV), and OM
covering a much wider energy range; furthermore, they model the foreground absorption with
{\tt TBabs}, an older version of {\tt TBnew} \citep{wil00}. On the other hand, the present
fits of the oxygen K region ($21{-}25$~\AA) with the {\tt warmabs} physical model yield a lower
column density ($N_{\rm H}=1.38\pm 0.1\times 10^{21}$ cm$^{-2}$) than that derived from our broadband fit
and {\tt TBnew}. This discrepancy is expected due to the different atomic data sets involved, particularly
since {\tt TBnew} only includes the photoabsorption cross section for neutral oxygen. Taking into
account the scatter of the previous column densities in the literature and the fact that we have adequately
modeled the absorption features from \ion{O}{1}, \ion{O}{2}, and \ion{O}{3}, we would expect
the value derived from {\tt warmabs} to be more reliable than that from {\tt TBnew}.


\section{Oxygen column density}\label{OxyGrow}

In order to derive the O column density along the line of sight of XTE~J1817-330, we have calculated the curve of growth for each of the oxygen K$\alpha$ transitions using the atomic data from \citet{gar05}, and equivalent widths (EWs) have been obtained using the functional fit described in Section~\ref{secGauss}. Table~\ref{grow1} shows a comparison between the present K$\alpha$ EWs for \ion{O}{1}, \ion{O}{2}, \ion{O}{3}, \ion{O}{6}, and \ion{O}{7} and those by \citet{sal07} for XTE J1817-330. EWs listed in \citet{jue04} for seven other sources are also included as well as those by \citet{yao09} for Cygnus X-2. For \ion{O}{1} K$\alpha$, our EW ($51\pm 5$~m\AA) is around the lower limits of \citet{sal07} and \citet{jue04}, and for \ion{O}{2} K$\alpha$, the present value of $52\pm 5$~m\AA\ is within the range that \citet{jue04} associated with the ISM. However, we find the K$\alpha$ lines in both \ion{O}{1} and \ion{O}{2} to be saturated which could influence our derived EWs. Furthermore, the K$\alpha$ EWs in \ion{O}{3}, \ion{O}{6}, and \ion{O}{7} are in good agreement with the values obtained by \citet{jue04}, \citet{sal07}, and \citet{yao09}.

Table~\ref{grow2} shows a comparison of the oxygen column densities obtained from
the EWs and the {\tt warmabs} model fit. As previously mentioned in Section~\ref{secGauss},
for the ionization parameter found in our best {\tt warmabs} fit, only \ion{O}{1}, \ion{O}{2}, and
\ion{O}{3} are included in the model. As a reference, we have also included the oxygen column density
for \ion{O}{6} from \citet{sav03}, obtained from FUSE data for four sources located near the same Galactic
latitude as XTE J1817-330. The curve-of-growth values for \ion{O}{1}, \ion{O}{2}, and \ion{O}{3} result in a total column density of $N_{\mathrm O}=N_\mathrm{O I}+N_\mathrm{O II}+N_\mathrm{O III}=5.71\times 10^{17}$~cm$^{-2}$, in relatively
good agreement with that derived from the ionic fractions in the {\tt warmabs} fit, namely
$N_{\mathrm O}=6.41 \times 10^{17}$~cm$^{-2}$.
However, the ratios \ion{O}{1}/\ion{O}{2} are significantly different: $1.4$ and $11.9$, respectively.
Due to line saturation, the measured EWs are likely to be underestimated, in particular 
for the \ion{O}{1} K$\alpha$ line.  This leads to a higher uncertainty in the column and ion fractions estimated from
the EW measurements. On the other hand, since the parameters in the {\tt warmabs} fit are constrained by all
the lines as well as the K edge, we are more confident about their reliability.
In the case of the \ion{O}{6} and \ion{O}{7} K$\alpha$ transitions, because the EWs are on the flat section of the curve of growth, they strongly depends on velocity dispersion. For \ion{O}{6} K$\alpha$, the oxygen column density spans from $N_\mathrm{O_{VI}}=71.12 \pm 29.15 \times 10^{15}$ cm$^{-2}$ to $N_\mathrm{O_{VI}}=5.06 \pm 2.53 \times 10^{15}$ cm$^{-2}$
when using dispersion velocities of $v=20$~km~s$^{-1}$ and $v=200$~km~s$^{-1}$, respectively. The smallest value is
still about one order of magnitude larger than that reported by \cite{sav03}, making difficult to constrain the 
the velocity dispersion. Also, no real improvement is achieved by considering larger dispersions, as the column density
is reduced very little once the velocities are larger than $200$~km~s$^{-1}$. This may be an indication that the observed \ion{O}{6} line is likely to originate from the neighborhood of XTE J1817-330 rather than from the ISM. In the case of \ion{O}{7}, the oxygen column density varies from $N_\mathrm{O_{VII}}=4.48\pm 1.84 \times 10^{18}$~cm$^{-2}$ ($v=20$~km~s$^{-1}$) to $N_\mathrm{O_{VII}}=4.5\pm 1.8 \times 10^{16}$ cm$^{-2}$ ($v=200$~km~s$^{-1}$).


\section{Discussion and conclusions}

There are several interesting conclusions emerging from this work that configure a set of useful guidelines in the analysis of ISM absorption features in {\em Chandra} and {\em XMM-Newton} spectra. In order to resolve such spectral signatures, bright sources with high count rates are desirable. In the present XTE~J1817-330 study case, we were fortunate to find four good quality spectra that were fitted simultaneously so as to obtain improved statistics. However, it was realized that the pileup effect must be taken into account in order to obtain reasonably reliable hydrogen column densities. This effect was treated adequately with the nonlinear pileup convolution model {\tt simple\_gpile2.sl}, whereby the pileup was estimated by fitting the continuum to be on average around 25\%. The resulting hydrogen column density is $N_{\rm H}=1.66^{+0.03}_{-0.04}\times 10^{21}$ cm$^{-2}$, which is somewhat ($\lesssim 7$\%) higher than those by \citet{kal05} and \citet{sal07}, the former obtained from 21~cm data.

The absorption features in the oxygen region (21--25~\AA) were first picked up with a functional model (power law and Gaussian profiles) which yielded K$\alpha$ lines of \ion{O}{1}, \ion{O}{2}, \ion{O}{3}, \ion{O}{6}, and \ion{O}{7} previously observed by \citet{sal07}, but for the first time, also K resonances of \ion{O}{1} and \ion{O}{2} with principal quantum number $n>2$. These line identifications at least confirm a two-phase plasma with both neutral (mostly) and highly ionized components. In the present work we have been mainly concerned with the former. In this respect,
since neutral oxygen is not the sole charge state responsible for the cold oxygen absorption, the four exposures are then simultaneously fitted with a representative physical model referred to as {\tt warmabs}, which contains the atomic data for the oxygen isonuclear sequence computed by \citet{gar05}. In this step, the hydrogen column density is held fixed at the previously determined value of $N_{\mathrm H}=1.66 \times 10^{21}$~cm$^{-2}$ while the ionization parameters and oxygen abundance are treated as free parameters. A detailed comparison of this model with observations brings forth the inaccuracies of the atomic data, which led us to introduce wavelengths shifts in both the K$\alpha$ line positions and photoabsorption cross sections of \ion{O}{1} and \ion{O}{2}.

The scope of this corrective procedure is severely limited by the general poor availability of laboratory data for the oxygen inner-shell, but it certainly leads to improved fits and statistics; in particular, to a well-constrained
ionization parameter of $\log\xi=-2.699\pm 0.023$ consistent with the absorption lines observed. In order to further improve the fit, the column density is allowed to vary resulting in a more reliable value ($N_{\mathrm H}=1.38\pm 0.1 \times 10^{21}$~cm$^{-2}$) which is within 13\% of previous estimates \citep{kal05, sal07}.
Furthermore, the fit yields an oxygen abundance of $A_{\mathrm O}= 0.689^{+0.015}_{-0.010}$ and ionization fractions of \ion{O}{1}/O = 0.911, \ion{O}{2}/O = 0.077 and \ion{O}{3}/O = 0.012. Regarding the oxygen abundance, it may be clarified that the quoted value is given relative to the solar standard ($8.83\pm0.06$) of \citet{gre98}. If it is rescaled to the recently revised standard by \citet{asp09} of $8.69\pm0.05$, then an ISM abundance ($A_{\mathrm O}= 0.952^{+0.020}_{-0.013}$) close to solar is obtained that is in excellent accord with the averaged value by \citet{bau06}, and can be included as further support for the solar abundance revision. Moreover, our ionization fractions are in good agreement with those found by \citet{jue04} of \ion{O}{2}/\ion{O}{1} $\approx 0.1$ and \ion{O}{3}/\ion{O}{1} $\leq 0.1$.

The functional fit allows us to derive oxygen column densities by means of a curve of growth method using the EWs for each of the observed K$\alpha$ absorption lines, which can then be compared with the values obtained from the {\tt warmabs} fit. Even though we find by both methods a relatively consistent total column density ($5.71\times 10^{17}$~cm$^{-2}$ in the curve of growth and $N=6.41 \times 10^{17}$~cm$^{-2}$ in
the {\tt warmabs} model fit), the \ion{O}{1}/\ion{O}{2} ratio for the K$\alpha$ lines is significantly lower (an order of magnitude) in the former. This is attributed to the high saturation level in both lines which leads to underestimates of the measured EWs, and we are therefore confident that the results
from the {\tt warmabs} model fit are more reliable. In the case of \ion{O}{6}, the width of the line strongly depends 
on the velocity dispersion of the gas. For the velocity dispersion values expected in the ISM ($v\sim20-200$~km~s$^{-1}$), 
the EWs obtained with our functional fit yield oxygen column densities far above from the FUSE values by \citet{sav03}. This may be an indication that the \ion{O}{6} observed absorption in the XTE~J1817-330 spectra is bound to occur near the source rather than in the ISM.

We find acceptable fits to the observed oxygen K-edge spectrum with a model consisting of atomic oxygen ions if the energies are slightly adjusted. However, it is also of great interest to detect or set limits on features due to other phases of interstellar oxygen and other elements, namely molecules or solids (dust). It is expected that these phases comprise a significant fraction of the total oxygen column in the ISM. Evidence for this includes observed depletions of atomic oxygen along many lines of sight relative to certain elements \citep{jen09, yao09, whi10}, together with observations of molecular features in other wavelength bands. An example of a potentially detectable signature of such material is the K resonance of the CO molecule expected at a rest wavelength of 23.21~\AA\ \citep{bar79}. This line is separated from the nearest feature we clearly detect by approximately 100~m\AA, which can be compared with the magnitude of the wavelength shifts we adopt in order to get acceptable fits, the largest of which is 33~m\AA.  Therefore, we conclude that the existing atomic absorption data are not sufficiently uncertain to mask the presence of this molecular signature. The search for molecular and solid features in X-ray spectra remains of great interest, both as tests of the atomic data and of the models for the ISM.

We believe that the present benchmark has produced an improved version of the {\tt warmabs} model which can now be used for a more extensive and reliable study of oxygen K photoabsorption towards other X-ray sources. This will be the next step of the current study where the oxygen abundance variations in the ISM are certainly of interest. In a similar fashion, we also intend to evaluate the Ne and Fe edge regions for ongoing physical model refinement.


\acknowledgments
Part of this work was carried out by Efra\'in Gatuzz during attendance in July 2011
to the Committee On Space Research (COSPAR) Capacity Building Workshop in San Juan,
Argentina, and visits in August 2011 to the Laboratory of High Energy
Astrophysics, NASA Goddard Space Flight Center, Greenbelt, Maryland, USA,
and in February--March 2012 to the European Space Astronomy Centre (ESAC),
Madrid, Spain, the latter funded by the COSPAR Fellowship Program. Warm hospitality
and tutelage at these institutions are kindly acknowledged, in particular from
Andy Pollock and Carlos Gabriel at ESAC. We would also like to thank Michael Nowak
from MIT for useful discussions relating the pileup model and the {\sc isis} implementation.

%
\bibliographystyle{apj}
\bibliography{my-references}
%
\clearpage
%
\clearpage
\clearpage
\clearpage
\clearpage
\clearpage

\clearpage

\begin{deluxetable}{lllll}
\tablecaption{Comparison of EBIT and theoretical wavelengths (\AA) for K lines \label{ebit}}
\tablewidth{0pt}
\tablehead{
\colhead{Ion} & \colhead{Lower level ($J$)} & \colhead{Upper level ($J'$)} & \colhead{EBIT} &  \colhead{Theory}
}
\startdata
\ion{O}{6} & ${\rm 1s^22s(1/2)}$ & ${\rm 1s2s2p(1/2, 3/2)}$ & $22.0194\pm 0.0016^a$ & 22.00$^c$ \\
           &                     &                          & $22.374\pm 0.008^b$    & 22.05$^d$ \\
           &                     &                          &                       & 22.03$^e$ \\
\ion{O}{5} & ${\rm 1s^22s^2(0)}$ & ${\rm 1s2s^22p(1)}$      & $22.374\pm 0.003^a$   & 22.33$^c$ \\
           &                     &                          & $22.370\pm 0.010^b$   & 22.35$^d$ \\
           &                     &                          &                       & 22.37$^e$ \\
\ion{O}{4} & ${\rm 1s^22s^22p(1/2, 3/2)}$ & ${\rm 1s2s^22p^2(1/2, 3/2)}$ & $22.741\pm 0.005^b$ & 22.78$^c$ \\
           &                     &                          &                                  & 22.73$^d$ \\
           &                     &                          &                                  & 22.75$^e$ \\
\ion{O}{3} & ${\rm 1s^22s^22p^2(1, 2)}$ & ${\rm 1s2s^22p^3(1)}$ & $22.071\pm 0.006^b$ & 23.08$^c$ \\
           &                     &                          &                         & 23.05$^d$ \\
           &                     &                          &                         & 23.07$^e$ \\
\enddata
\tablenotetext{a}{EBIT measurement \citep{sch04b}}
\tablenotetext{b}{EBIT measurement \citep{gu05}}
\tablenotetext{c}{{\sc hullac} calculation \citep{beh02}}
\tablenotetext{d}{$R$-matrix calculation \citep{pra03}}
\tablenotetext{e}{{\sc hfr} calculation \citep{gar05}}
\end{deluxetable}

\clearpage
\begin{deluxetable}{cccccc}
\tablecaption{{\em Chandra} observations used in this paper \label{tab1}}
\tablewidth{0pt}
\tablehead{
\colhead{ObsID} & \colhead{Date} & \colhead{Exposure (ks)} & \colhead{Instrument} & \colhead{Grating} &
\colhead{Label}
}
\startdata
6615&2006 Feb 13&18 & HETG-ACIS &MEG&Observation 1\\
6616&2006 Feb 24&29 & HETG-ACIS&MEG&Observation 2\\
6617&2006 Mar 15&47 & HETG-ACIS&MEG&Observation 3\\
6618&2006 May 22&51 & HETG-ACIS&MEG&Observation 4\\
\enddata
\end{deluxetable}


\clearpage
 \begin{deluxetable}{lll}
\tabletypesize{\scriptsize}
 \tablecaption{Broadband simultaneous fit parameters \label{tab2}}
\tablewidth{0pt}
\tablehead{
\colhead{Model} & \colhead{Parameter} & \colhead{Value}
}
\startdata
TBNew &$N_{\rm H}$ ($10^{21}$~cm$^{-2}$)&$ 1.66^{+0.03}_{-0.04}$ \\
TBNew &Neon abundance (A$_{\rm Ne}$)&$  2.258\pm 0.080     $ \\
TBNew &Iron abundance (A$_{\rm Fe}$)&$  1.302\pm 0.050     $ \\
TBNew &Oxygen abundance (A$_{\rm O}$)&$ 1.178\pm 0.022     $ \\
Power-law &Normalization &$ 9.416\pm 0.064   ^{a}$ \\
                       & &$ 7.039\pm 0.041   ^{b}$ \\
                       & &$ 4.613\pm 0.042   ^{c}$ \\
                       & &$ 1.481\pm 0.001   ^{d}$ \\
Power-law &Photon Index&$ 1.545\pm 0.030     ^{a}$ \\
                       & &$ 1.618\pm 0.030   ^{b}$ \\
                       & &$ 1.735\pm 0.029   ^{c}$ \\
                       & &$ 2.242\pm 0.038   ^{d}$ \\
SimpleGpile2 &$\beta$  &$ 0.050\pm 0.001     ^{a}$ \\
                       & &$ 0.051\pm 0.001   ^{b}$ \\
                       & &$ 0.050\pm 0.001   ^{c}$ \\
                       & &$ 0.049\pm 0.001   ^{d}$ \\
Reduced chi-square     & $\chi^{2}$ &$ 1.208   $\\
\enddata
\tablecomments{Abundances relative to the solar values of \citet{wil00}.}
\tablenotetext{a}{ObsID 6615}
\tablenotetext{b}{ObsID 6616}
\tablenotetext{c}{ObsID 6617}
\tablenotetext{d}{ObsID 6618}
\end{deluxetable}

%
%

\clearpage
\begin{deluxetable}{llcccc}
 \tablecaption{Absorption line assignments \label{tab4}}
\tablewidth{0pt}
\tablehead{
\colhead{Ion} & \colhead{Transition} & \colhead{$\lambda^a$ (\AA)}& \colhead{$\lambda^b$ (\AA)}
}
\startdata
\ion{O}{7} & ${\rm 1s}^2\ ^1{\rm S} - [{\rm 1s}]{\rm 2p}\ ^1{\rm P^o}$                       & $21.589\pm 0.003$ & $21.60\pm 0.06$ \\
\ion{O}{6} & ${\rm 2s}\ ^2{\rm S} - [{\rm 1s}]{\rm 2s2p}\ ^2{\rm P^o}$                       & $22.022\pm 0.003$ &                 \\
\ion{O}{2} & ${\rm 2p}^3\ ^4{\rm S^o} - [{\rm 1s}]{\rm 2p}^3{\rm 4p}\ ^4{\rm P}$             & $22.101\pm 0.005$ &                 \\
\ion{O}{2} & ${\rm 2p}^3\ ^4{\rm S^o} - [{\rm 1s}]{\rm 2p}^3{\rm 3p}\ ^4{\rm P}$             & $22.280\pm 0.003$ &                 \\
\ion{O}{1} & ${\rm 2p}^4\ ^3{\rm P} - [{\rm 1s}]{\rm 2p}^4(^2{\rm P})4{\rm p}\ ^3{\rm P^o}$  & $22.609\pm 0.007$ &                 \\
\ion{O}{1} & ${\rm 2p}^4\ ^3{\rm P} - [{\rm 1s}]{\rm 2p}^4(^2{\rm P})3{\rm p}\ ^3{\rm P^o}$  & $22.686\pm 0.004$ &                 \\
\ion{O}{1} & ${\rm 2p}^4\ ^3{\rm P} - [{\rm 1s}]{\rm 2p}^4(^4{\rm P})4{\rm p}\ ^3{\rm P^o}$  & $22.790\pm 0.001$ &                 \\
\ion{O}{1} & ${\rm 2p}^4\ ^3{\rm P} - [{\rm 1s}]{\rm 2p}^4(^4{\rm P})3{\rm p}\ ^3{\rm P^o}$  & $22.884\pm 0.004$ & $22.91\pm 0.03$ \\
\ion{O}{3} & ${\rm 2p}^2\ ^3{\rm P} - [{\rm 1s}]{\rm 2p}^3\ ^3{\rm P^o}$ & $23.054\pm 0.001$ & $23.13\pm 0.09$ \\
\ion{O}{3} & ${\rm 2p}^2\ ^3{\rm P} - [{\rm 1s}]{\rm 2p}^3\ ^3{\rm D^o},\ ^3{\rm S^o}$       & $23.104\pm 0.005$ & $23.13\pm 0.09$ \\
\ion{O}{2} & ${\rm 2p}^3\ ^4{\rm S^o} - [{\rm 1s}]{\rm 2p}^4\ ^4{\rm P}$ & $23.358\pm 0.002$ & $23.35\pm 0.03$ \\
\ion{O}{1} & ${\rm 2p}^4\ ^3{\rm P} - [{\rm 1s}]{\rm 2p}^5\ ^3{\rm P^o}$ & $23.502\pm 0.001$ & $23.52\pm 0.02$ \\
\enddata
\tablenotetext{a}{Present work}
\tablenotetext{b}{\citet{sal07}}
\end{deluxetable}

\clearpage
\begin{deluxetable}{lll}
 \tabletypesize{\scriptsize}
 \tablecaption{{\tt warmabs} simultaneous fit parameters\label{tab5}}
\tablewidth{0pt}
\tablehead{
\colhead{Model} & \colhead{Parameter} & \colhead{Value}
}
\startdata
Warmabs            & $N_{\rm H}$ ($10^{21}$~cm$^{-2}$)    & $1.38\pm 0.01$              \\
Warmabs            & Log ionization parameter (log $\xi$) & $-2.699\pm0.023$            \\
Warmabs            & Oxygen abundance (A$_{\rm O}$)       & $0.689^{+0.015}_{-0.010}$   \\
Power-law          & Normalization                        & $2.384^{a}$ \\
                   &                                      & $2.544^{b}$ \\
                   &                                      & $1.511^{c}$ \\
                   &                                      & $0.685^{d}$ \\
Power-law          & Photon Index                         & $3.044^{a}$ \\
                   &                                      & $2.479^{b}$ \\
                   &                                      & $2.720^{c}$ \\
                   &                                      & $2.617^{d}$ \\
Reduced chi-square & $\chi^{2}$                           & $1.245$     \\
\enddata
\tablecomments{Oxygen abundance relative to the solar value of \citet{gre98}.}
\tablenotetext{a}{ObsID 6615}
\tablenotetext{b}{ObsID 6616}
\tablenotetext{c}{ObsID 6617}
\tablenotetext{d}{ObsID 6618}
\end{deluxetable}

\clearpage
\begin{deluxetable}{ll}
\tabletypesize{\scriptsize}
 \tablecaption{Hydrogen column density comparison \label{tab3}}
\tablewidth{0pt}
\tablehead{
 \colhead{Method} & \colhead{N$_H$ ($10^{21}$~cm$^{-2}$)}
}
\startdata
{\tt TBnew} fit$^a$ & $1.66^{+0.03}_{-0.04}$ \\
{\tt warmabs} fit$^a$ & $1.38\pm 0.1$ \\
$21$~cm survey$^b$   & 1.58                   \\
$21$~cm survey$^c$   & 1.39                   \\
Spectral model fit$^d$   & $1.55\pm 0.05$         \\
Spectral model fit$^{e}$   & 1.2                    \\
Spectral model fit$^f$   & $0.88 - 0.97$         \\
Spectral model fit$^g$   & $1.0 - 3.0$         \\
Spectral model fit$^h$   & $1.08\pm 0.05$         \\
\enddata
\tablenotetext{a}{Present work}
\tablenotetext{b}{\citet{dic90}}
\tablenotetext{c}{\citet{kal05}}
\tablenotetext{d}{\citet{sal07}}
\tablenotetext{e}{\citet{ryk07}, \citet{gie08}, \citet{roy11}}
\tablenotetext{f}{\citet{mil06b}}
\tablenotetext{g}{\citet{tor06}}
\tablenotetext{h}{\citet{gie09}}
\end{deluxetable}

\clearpage
\begin{deluxetable}{llllll}
\tabletypesize{\scriptsize}
 \tablecaption{Equivalent widths comparison for the oxygen K$\alpha$ transitions in the ISM \label{grow1}}
\tablewidth{0pt}
\tablehead{
  \colhead{Source } &\colhead{O I K$\alpha$} &\colhead{O II K$\alpha$} & \colhead{O III K$\alpha$}& \colhead{O VI K$\alpha$}& \colhead{O VII K$\alpha$}\\
  &(m\AA)&(m\AA)&(m\AA)&(m\AA)&(m\AA)
}
\startdata
XTE J1817-330 $^{a}$&$51\pm	5$ &$52\pm	 5$  &$16\pm 5$ &$14\pm 5$& $ 54\pm	4$  \\
XTE J1817-330 $^{b}$&$70  \pm 20 $ &$40  \pm 20 $  &$30  \pm 30 $ && $40  \pm 30 $   \\
Cyg X-2 $^{c}$&67$\pm$11 &26$\pm$17  &9$\pm$8   &&  \\
Cyg X-2 $^{d}$& $50\pm 3$ &$28\pm 6$  & & $11\pm 2$&  \\
4U 1543-624 $^{c}$&68$\pm$11 &41$\pm$11   &18$\pm$5 &&   \\
4U 1820-30 $^{c}$&70$\pm$20 &41$\pm$17  & 20$\pm$10 &&   \\
4U 1735-44 $^{c}$&80$\pm$30 &40$\pm$20  & 24$\pm$10 &&   \\
GX 9+9 $^{c}$&80$\pm$30 &50$\pm$19  & 27$\pm$13 &&  \\
4U 1636-53 $^{c}$&90$\pm$30 &60$\pm$20  & 33$\pm$13 &&  \\
Cyg X-1 (ObsID 3407) $^{c}$&95$\pm$17 &20$\pm$11  & 35$\pm$15  &&  \\
\enddata
\tablenotetext{a}{Present work}
\tablenotetext{b}{\citet{sal07}}
\tablenotetext{c}{\citet{jue04}}
\tablenotetext{d}{\citet{yao09}}
\end{deluxetable}

\clearpage
\begin{deluxetable}{lccccc}
\tabletypesize{\scriptsize}
 \tablecaption{Oxygen column density comparison \label{grow2}}
\tablewidth{0pt}
\tablehead{
  \colhead{Source } &\colhead{O I   } &\colhead{O II   } & \colhead{O III   }& \colhead{O VI  }& \colhead{O VII  }\\
  &($10^{17}$~cm$^{-2}$)&($10^{17}$~cm$^{-2}$)&($10^{17}$~cm$^{-2}$)&($10^{15}$~cm$^{-2}$)&($10^{15}$~cm$^{-2}$)
}
\startdata
 XTE J1817-330 $^{a}$&   $3.18 \pm 0.79$     &$2.18 \pm 0.54$  & $0.35 \pm 0.08$ & $5.06 \pm 2.53$ $^{d}$ &$45\pm 18$ $^{d}$ \\
 &&&& $ 5.99 \pm 1.9$ $^{e}$  &$988 \pm 405$ $^{e}$ \\
  &&&&$71.12 \pm 29.15$ $^{f}$& $ 4484 \pm 1838 $ $^{f}$  \\
  XTE J1817-330 $^{b}$&$5.85 \pm 1.75$        &$0.49 \pm 0.12$  &$0.07\pm 0.02$\\
 PG 1302-102 $^{c}$& &   && $0.16\pm 0.03$    \\
 Mrk 1383 $^{c}$& &  & & $0.38 \pm 0.07$    \\
 ESO 141-G55 $^{c}$& & & &  $0.31\pm 0.06$     \\
 PKS 2005-489 $^{c}$& & & &  $0.60\pm 0.01 $    \\
\enddata
\tablenotetext{a}{Calculated from the EWs and the curves of growth (see section \ref{OxyGrow} ). }
\tablenotetext{b}{Derivated from the {\tt warmabs} model fit using the solar value of \citet{gre98} (see section \ref{secGauss}) }
\tablenotetext{c}{Obtained from FUSE data by \citet{sav03}}
\tablenotetext{d}{Using a velocity dispersion $v=200$ km/s}
\tablenotetext{e}{Using a velocity dispersion $v=100$ km/s}
\tablenotetext{f}{Using a velocity dispersion $v=20$ km/s}
\end{deluxetable}

\end{document}